\documentclass[preprint2]{aastex}
\usepackage{graphicx} 
\usepackage{verbatim}
\usepackage{natbib}
\usepackage{amssymb}
\usepackage{amsmath}
\usepackage{times}
\usepackage{pstricks}
\usepackage{epsfig}
\usepackage{enumerate}

\def\ls{\mathrel{\raise0.27ex\hbox{$<$}\kern-0.70em 
\lower0.71ex\hbox{{$\scriptstyle \sim$}}}}
\def\gs{\mathrel{\raise0.27ex\hbox{$>$}\kern-0.70em 
\lower0.71ex\hbox{{$\scriptstyle \sim$}}}}

\newcommand{\jpl}{1}
\newcommand{\nwu}{2}
\newcommand{\edi}{3}
\newcommand{\har}{4}
\newcommand{\cali}{5}
\newcommand{\shear}{\mbox{$\mathcal{S}$}}

\begin{document}

\title{Astronomical image simulation for telescope and survey development}

\renewcommand{\thefootnote}{\fnsymbol{footnote}}
\author{
Benjamin M.\ Dobke \altaffilmark{\jpl},
David E.\ Johnston \altaffilmark{\nwu},
Richard Massey \altaffilmark{\edi},
F.\ William High \altaffilmark{\har},
Matt Ferry \altaffilmark{\cali},
Jason Rhodes \altaffilmark{\jpl,\cali},
R.\ Ali Vanderveld \altaffilmark{\jpl, \cali}
}

\altaffiltext{\jpl}{Jet Propulsion Laboratory, California Institute of Technology \\ 4800 Oak Grove Drive, Pasadena, CA 91109, United States}
\altaffiltext{\nwu}{Northwestern University, Department of Physics and Astronomy 2145 Sheridan Rd, Evanston, IL 60208, United States}
\altaffiltext{\edi}{Institute for Astronomy, University of Edinburgh, Royal Observatory  Blackford Hill, Edinburgh, EH9 3HJ, United Kingdom}
\altaffiltext{\har}{Harvard University, Department of Physics, 17 Oxford St. \\ Cambridge, MA 02138, United States}
\altaffiltext{\cali}{California Institute of Technology, 1201 E. California Blvd. \\ Pasadena, CA 91125, United States}

\begin{abstract}

We present the \emph{simage} software suite for the simulation of artificial extragalactic images, based empirically around real observations of the Hubble Ultra Deep Field (UDF).  The simulations reproduce galaxies with realistic and complex morphologies via the modeling of UDF galaxies as \emph{shapelets}.  Images can be created in the \emph{B, V, i} and \emph{z} bands for both space- and ground-based telescopes and instruments. The simulated images can be produced for any required field size, exposure time, Point Spread Function (PSF), telescope mirror size, pixel resolution, field star density, and a variety of detector noise sources.  It has the capability to create images with both a pre-determined number of galaxies or one calibrated to the number counts of pre-existing data sets such as the HST COSMOS survey.  In addition, simple options are included to add a known weak gravitational lensing (both shear and flexion) to the simulated images.  The software is available in Interactive Data Language (IDL) and can be freely downloaded for scientific, developmental and teaching purposes.

\end{abstract}


\section{Introduction}
\label{sect:intro} 

In the next decade, the quantity of data available to cosmology will rapidly increase.  New telescopes, both on the ground
and in space, promise to image many thousands of square degrees.  The cosmology community is now tasked with developing
methods to analyze such data, and to extract as much information from various astronomical phenomena.  These methods need to achieve unprecedented precision if the promise (and potential statistical power) of future  surveys are to be fully exploited.

Developing image analysis tools requires realistic mock data, containing as many instrumental effects as possible, plus a
known, underlying cosmological signal, against which measurements can be judged.  To generate galaxy shapes in simulated
ground-based images, \emph{Skymaker} \citep{erbe} uses a simple physical model of concentric isophotes with a de
Vaucouleurs profile for elliptical galaxies, and an additional, exponential component for spirals.  By varying the model
parameters, one can generate an unlimited number of unique simulated galaxies.  However, deep field images from
space-based telescopes contain galaxies with features more complex than these smooth analytical models can reproduce. 
Here we present the full \emph{simage} pipeline (as gradually developed in \citealt{simage1}, \citealt{step2} and
\citealt{ferry}), which empirically mimics the complex morphologies of galaxies seen in real data, such as the
Hubble Ultra Deep Field (UDF: \citealt{beck}) or Hubble Space Telescope (HST) COSMOS survey \citep{scov}.  The galaxy morphologies are captured via a \emph{shapelet} decomposition
\citep{refr,refr2,bern,mass,mass2}, which also makes it easy to introduce a specified weak gravitational lensing signal,
useful to hone shear measurement methods.  An example of \emph{shapelet} galaxy image simulation includes \emph{Skylens} \citep{Mene1}.

The \emph{simage} code is written in the Interactive Data Language (IDL) and can be downloaded from
\url{http://www.astro.caltech.edu/\~rjm/shapelets}. It also requires the core \emph{shapelets} package, available from the
same location, and certain routines from the NASA Goddard Space Flight Center (GSFC) IDL astronomy user's library. Those required routines are bundled in the \emph{simage} download, but updated versions may be periodically available from \url{http://idlastro.gsfc.nasa.gov/}. In addition, exploiting the full potential of some sections of \emph{simage} require \emph{SExtractor}, available from \url{http://astromatic.iap.fr/ software/sextractor/}.

This paper highlights the full capabilities of the code, its general structure, and the a number of possible uses.  In section 2, we briefly detail previous applications and associated results, plus the strengths and weaknesses of the utilized shapelet formalism.  In section 3, we describe the structure of the software package, and discuss the main modules of the software and their uses.  In section 4, we exploit the software to investigate telescope survey depth.  We conclude in section 5.

\section{Capabilities and applications of {\textit{`simage'}}}
\label{sect:capabilities}

At the heart of \emph{simage} is the ability of the \emph{shapelets} method to efficiently and flexibly reconstruct
complex galaxy morphologies.  Shapelets are a complete, orthogonal set of basis functions, and a weighted linear
combination of these can represent any localised image (\citealt{refr}; \citealt{bern} (BJ02); \citealt{mass}).  This is analogous to a Fourier transform, where weighted combinations of sines and cosines can be used to reconstruct non-localised images.  The mathematical properties of the shapelet basis set make it particularly convenient for astronomical image processing, including quick convolutions with
Point Spread Functions (PSF), pixillation, and operations such as translations, rotations, magnifications and shears, that can
be used to add a known signal into a simulated image \citep{refr2, mass2}.  The ability to represent shears in a simple manner, allows for the inclusion of distorting shears produced by both the telescope's optical systems and weak gravitational lensing within galaxy clusters.  While a Gaussian-based shapelet representation of a galaxy with a particularly strong central peak, or extended wings, is not ideal due to the difficulty reproducing such features with Gaussian forms, a number of tests to reproduce exponential radial profiles with shapelet basis functions have show good reproduction of the majority of galaxy shapes with very little bias (\citealt{step2}).

The simulated images are capable of being produced in the \emph{B, V, i} and \emph{z} bands, since they are based on a galaxy morphology catalog pre-constructed from the Hubble UDF.  Hence, the available passbands allowed by the F485W, F606W, F7775W, and F850LP filters of the HST.

\begin{figure}[t]
\centering
\epsfig{figure=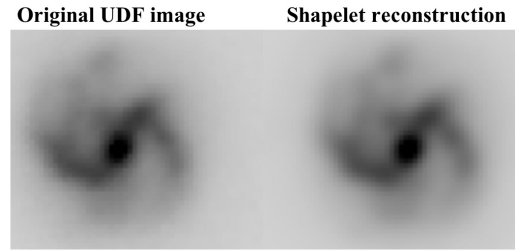,width=75mm} 
\caption{An example of a spiral galaxy (from the Hubble Deep Field) modeled using shapelets \citep{mass}.}  
\label{fig:one}
\end{figure}

\begin{figure*}[t]
\centering
\epsfig{figure=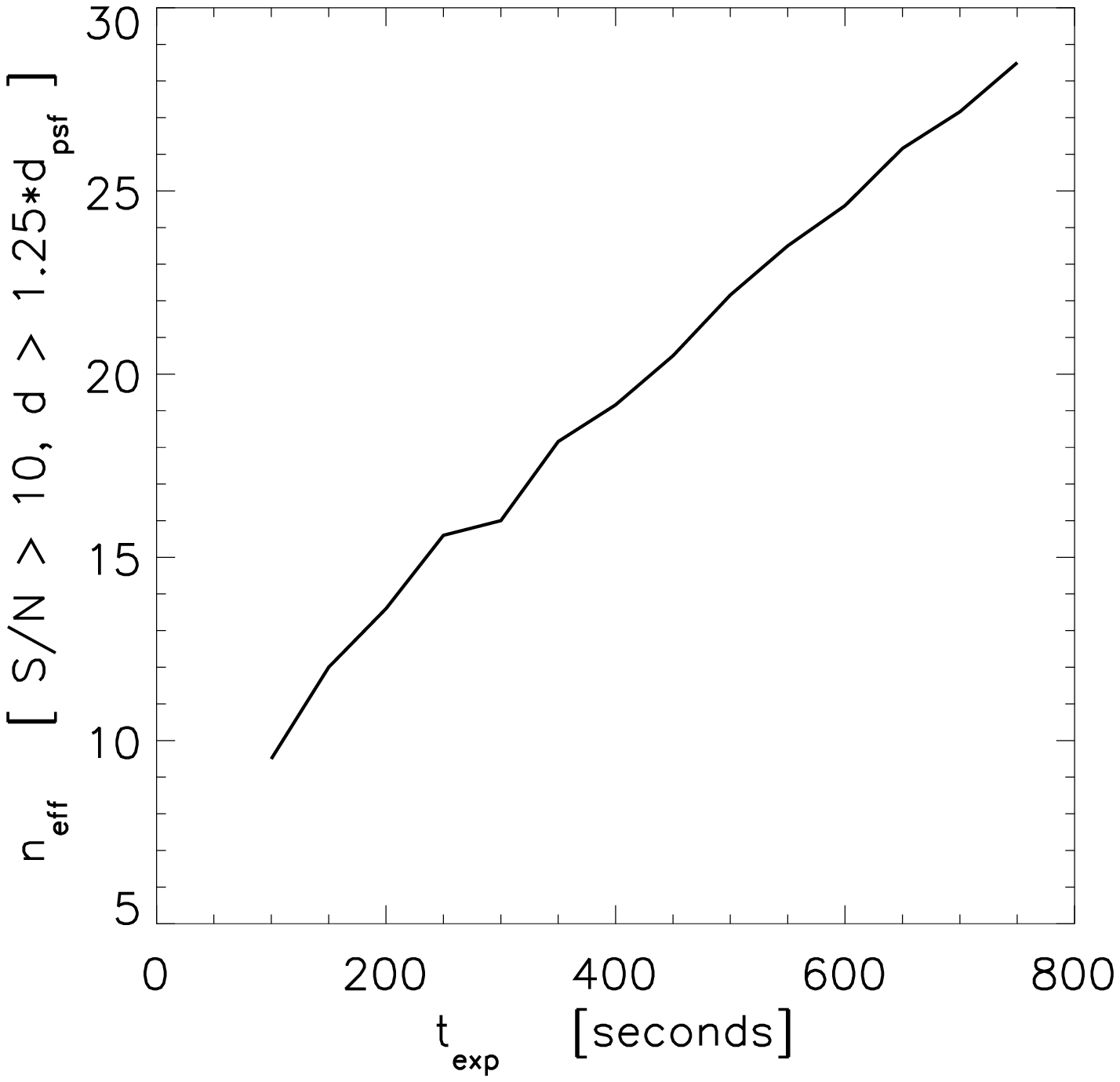,width=70mm} 
\epsfig{figure=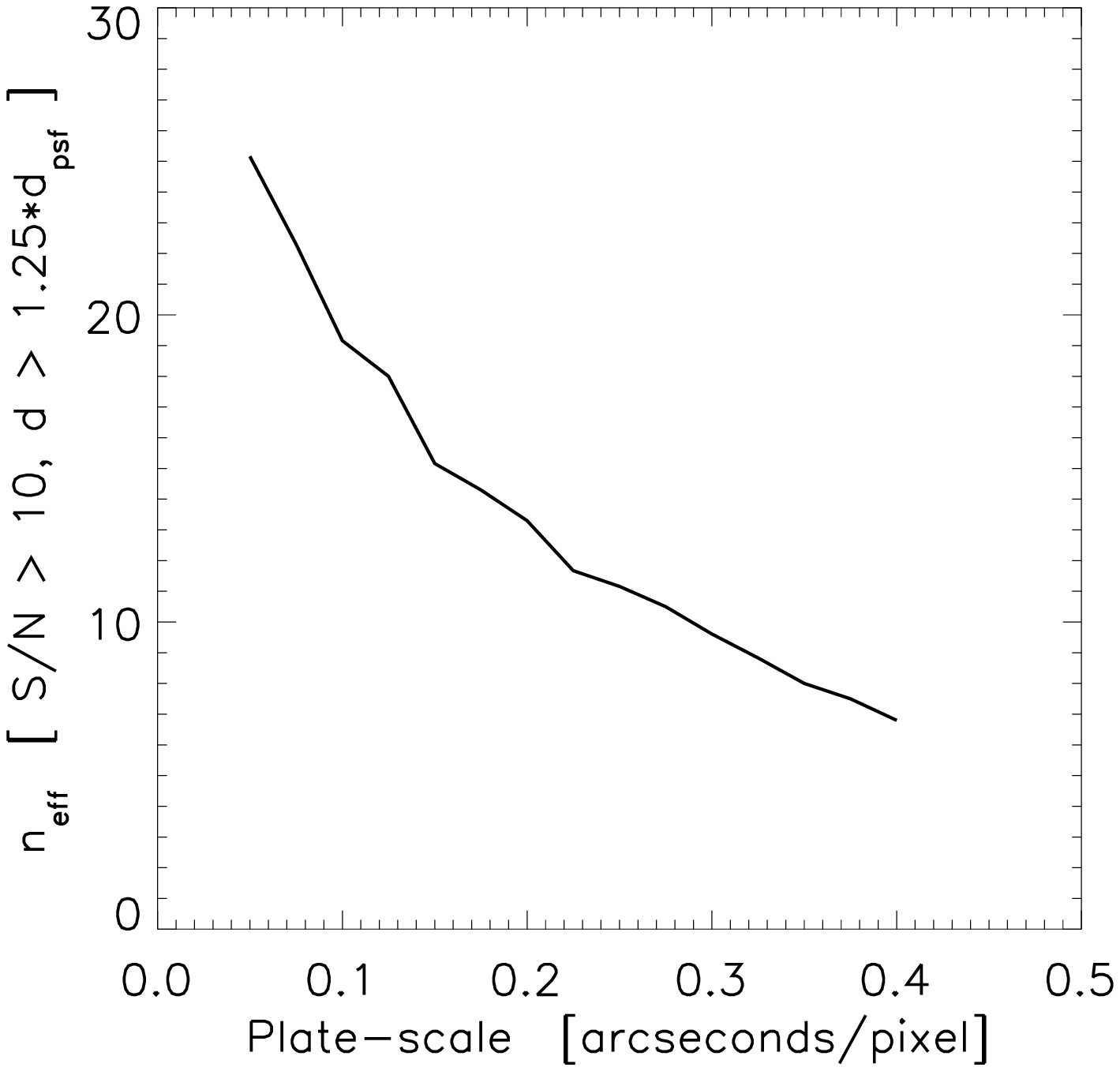,width=70mm}
\caption{
\emph{Left:} Plot of n$_{eff}$ galaxy counts vs. exposure time at a pixel scale of 0.1 arcseconds.  The $n_{eff}$ quantity in this context is the number of galaxies within a given survey area that meet a particular desired size and signal-to-noise ratio.  \emph{Right:} Plot of n$_{eff}$ galaxy counts vs. pixel scale at an exposure time of 400 seconds.  These particular plots were generated from images created in the \emph{i}-band (band 2 in \emph{simage}) for a mock-up telescope 1.2m diameter mirror over a 6 arcminute$^{2}$ area of the sky - see Table 1.
}
\label{fig:four}
\end{figure*}

For a more in depth description of the method, including a general mathematical introduction to its application to image simulation, the reader is directed to the aforementioned references and Appendix A.  In this section, we shall highlight the capabilities such a shapelet formalism provides, along with associated applications and results.

\subsection{Prior applications: Shear studies, data codec, and mission development}
\label{sect:apps}

Previous applications of the shapelets image simulation, in it's \emph{simage} incarnation as presented here, or in other forms, have ranged from direct image creation for community shear-analysis studies, data compression investigations, and telescope/instrumental development.  We will briefly detail these below.

The Shear TEsting Program (STEP) was a collaborative project which aimed to improve the accuracy and reliability of all weak gravitational lensing measurements in preparation for the next generation of wide-field surveys. STEP was launched in order to test and improve the accuracy and reliability of all these methods through the rigorous testing of shear measurement pipelines, the exchange of data and the sharing of technical and theoretical knowledge within the weak lensing community (\citealt{step2}).  Here, the shapelets based \emph{simage} code was used to create image data with incorporated shear fields for analysis in the testing program.  Subroutines of the code were also used in the analysis of the resultant data.  \citep{schr} used images with simulated shear produced by this software to verify and validate their weak lensing code, which they then used to provide the first direct detection of dark energy using weak lensing tomography.  Opening up the issue to wider participation, particularly computer scientists, the GRavitational lEnsing Accuracy Testing 2008 (GREAT08; \citealt{brid}; \citealt{brid2}), and again allowed for the application of \emph{simage}'s image simulation capabilities.

In \cite{vand}, \emph{simage} was utilised to test compression-decompression (codec) algorithms and methods for future visible survey telescopes.  This is of vital importance given the vast quantities of data both space- and ground-based survey telescopes are predicted to produce in the coming years (~10s of petabytes; e.g. LSST; \citealt{ivez}).  The roll of \emph{simage} here was to create batches of simulated image data that recreated sizes, morphology, fluxes, and shapes that accurately mimic those we might expect from a visible survey telescope in both orbit and on Earth.

Specific mission development has also been an application of the \emph{simage} routines.  In particular, the simulation pipeline has been a critical optimization tool in the optical design and observation strategy for ESA's cosmic visions candidate Euclid (\citealt{refr3}), Nasa/DOE Joint Dark Energy Mission (JDEM) concepts (e.g. SNAP; \citealt{jeli}; \citealt{high}), and the weak lensing balloon missions.
\subsection{Example application}
\label{sect:example}

As a demonstration of one of \emph{simage}'s potential uses, we present an example investigation intended to explore telescope survey depth, specifically the relation between the effective number of galaxies observed in a survey sample, $n_{eff}$, vs. pixel scale and exposure time for a mock-up space telescope design (the $n_{eff}$ quantity in this context is the number of galaxies within a given survey area that meet a particular desired size and signal-to-noise ratio).  A list of possible input telescope and instrument parameters are shown in Table 1.  We generate images according to these inputs using a varying the pixel scale between 0.05 and 0.40 arcseconds/pixel (with constant exposure = 400s), and the exposure time between 100-800 seconds (with constant pixel scale = 0.1"/pixel).  An example of the output as created from a set of images simulated from \emph{simage} is shown in Fig. \ref{fig:four}

\begin{table}[t]
\caption{Demonstration inputs within \emph{telescope.param} for a mock-up telescope design}
{\scriptsize 
\begin{center}
\begin{tabular}{|l|l|}
\hline
Parameter file input: & Value: 
\\\hline
\texttt{throughput\_ratio} & [0.0, 0.0, 1.2, 0.0, 0.0, 0.0]\\
\texttt{pixel\_scale} & [0.0, 0.0, 0.1, 0.0, 0.0, 0.0] arcsec./pixel\\
\texttt{read\_noise} & 5 electrons\\
\texttt{collecting\_area} & 1.13 m$^{2}$\\
\texttt{band\_begin} & 2\\
\texttt{band\_end} & 2\\
\texttt{exposure\_time} & 100 seconds\\
\texttt{area} & 6.0 square arcminutes\\
\texttt{n\_star} & 1000\\
\texttt{n\_gal} & Default\\
\texttt{ee50} & 0.15 arcseconds\\
\hline 
\end{tabular}
\end{center}
\label{default}
}
\end{table}

\section{The simulation pipeline}
\label{sect:pipeline}

This section will detail the critical internal routines of \emph{simage}, and describe the flow of the simulated image creation.

\subsection{Module overview}
\label{sect:modules}

The routine {\bf simage.pro} is the primary program; it utilizes various routines within the pipeline to manufacture a simulated image.  Keywords listed in Table~2 can be used to specify the desired telescope and survey characteristics.  By default, the image will be produced in the \emph{B, V, i} and \emph{z} bands, based on a galaxy morphology catalog pre-constructed from the Hubble UDF.  More permanent changes to the telescope and survey characteristics can be fixed in the {\bf telescope.param} file.

\begin{figure}
\centering
\epsfig{figure=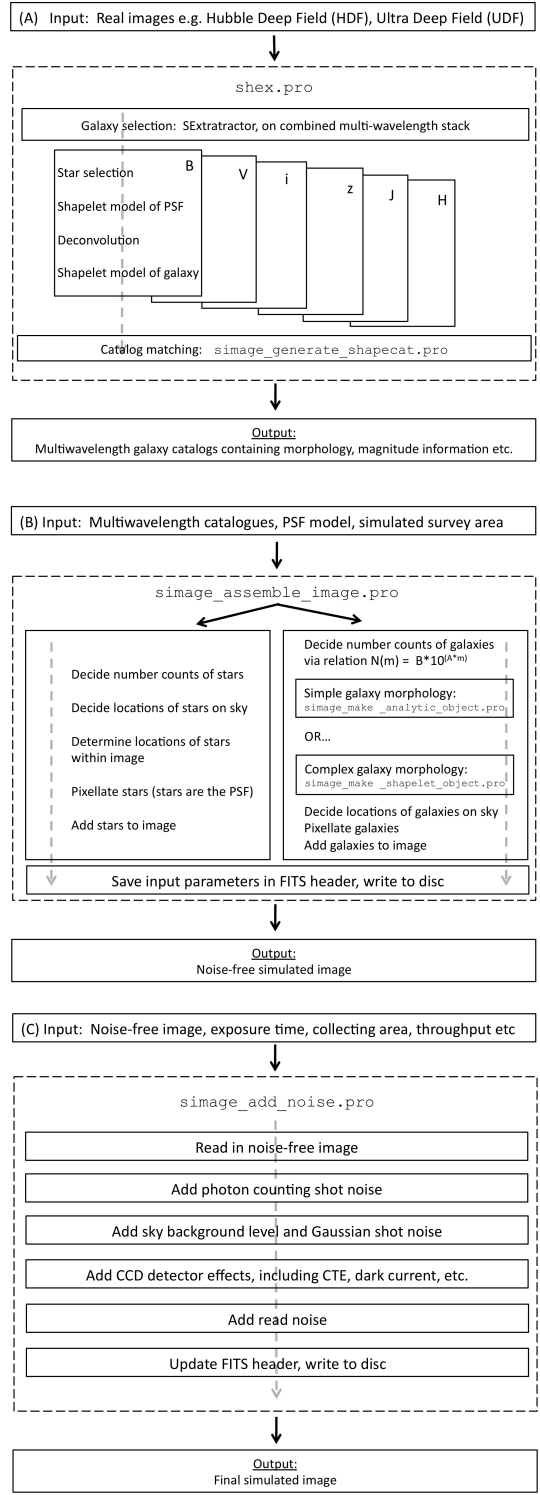,width=70mm} 
\caption{Flow chart of the simulation pipeline's key processes beginning with the initial input UDF images to the final output \emph{.\,fits} images.}  
\label{fig:two}
\end{figure}

Figure \ref{fig:two} details the pipeline's main processes in the form of a flow chart.  We note from the chart that there are three main stages to the pipeline;  the multi-wavelength catalog generation, the repopulation of the catalog images into a field/resolution governed by the desired telescope parameters, and finally the addition of the various noise components.  Other important routines in the pipeline are,

\begin{itemize}
\item {\bf simage\_make\_shapelet\_object.pro:}  Generates a pixellated image of one simulated galaxy from a given a set of shapelet coefficients.
\item {\bf simage\_make\_analytic\_object.pro:}  Generates an object for the image simulations, using an analytic profile.  The size, magnitude and ellipticity are drawn from a real UDF galaxy template.
\item {\bf num\_counts\_frac.pro} Calculates the galaxy magnitude distributions normalized to the COSMOS survey data at mid magnitudes \citep{Leau}, and to a compilation of other surveys at low and high magnitudes (\citealt{Metc}; see \S3.2.2 for discussion.)
\item {\bf get\_telescope\_psf*:}  Reads in the desired PSF \emph{fits} file before converting it into shapelet space (* `telescope' represents a variety of telescope or survey names included in the pipeline e.g. \emph{get\_udf\_psf}, etc.).  The PSF is an array of odd dimensions and be in logarithmic units, by convention.
\end{itemize}
 
The desired size of the PSF is quantified in two ways: via the energy that encloses 50\% of the PSF energy (EE50), or the Full-Width at Half Maximum (FWHM).  The EE50 term is more commonly used in optical engineering, whereas FWHM is more commonly used in science applications.  Both are available as user inputs and both have particular advantages when creating simulated images. For example, when comparing the EE50 to that of the FWHM, the former is more affected by the tail's profile.  For this reason, EE50 is a better measure of how compact something is if you only have one number to describe an object.  As a comparison, a Gaussian with a FWHM of 1.77 pixels, would have an EE50 of 0.885 pixels, while a typical PSF with the same FWHM of 1.77 pixels, would have an EE50 of 1.21 pixels.

\begin{table}
\caption{Descriptions of the user inputs for \emph{telescope.param}.  These parameter inputs govern telescope design and in turn the resultant output images.  Some parameter inputs are degenerate, e.g. if a \emph{throughput\_ratio} is entered, the \emph{filter\_files} path is ignored.}
{\scriptsize 
\begin{center}
\begin{tabular}{|l|l|}
\hline
Parameter file input: & Description: 
\\\hline
\texttt{throughput\_ratio} & Total system throughputs relative to UDF\\
\texttt{sky\_level} & Skylevel for each band (counts/s/arcsec$^{2}$) \\
\texttt{zeropoint} & Chosen zeropoint for the images in each band\\
\texttt{pixel\_scale} & The instrument pixel scale in arcsecond/pixel\\
\texttt{psf\_type} & Selects which PSF to use\\
\texttt{psf\_path} & Path to the user's chosen PSF .fits file\\
\texttt{read\_noise} & CCD read noise in number of electrons\\
\texttt{background} & 0 = background subtracted, 1 = added \\
\texttt{collecting\_area} & The mirror collecting area in m$^{2}$\\
\texttt{band\_begin} & The band on which to start the simulations\\
\texttt{band\_end} & The band on which to end the simulations\\
\texttt{exposure\_time} & Exposure time in seconds\\
\texttt{area} & The area on the sky to simulate in sq. arcmins\\
\texttt{random\_seed} & A random seed for all random selections\\
\texttt{gamma} & The user specified weak lensing shear\\
\texttt{output\_file\_pref} & Selection of output image file names\\
\texttt{n\_star} & Number of field stars to be added\\
\texttt{n\_gal} & Number of field galaxies\\
\texttt{filter\_files} & Path to user's transition filter files\\
\texttt{ee50} & The half light radius of the PSF\\
\texttt{fwhm} & The full-width at half-maximum of the PSF\\
\hline 
\end{tabular}
\end{center}
\label{default}
}
\end{table}

\begin{table}
\caption{Meaning of flags output from \emph{shex.pro} detailing to the user how well a given object was modeled with shapelets.  A 0 value implies a successful decomposition, while a 10 signals failure.  A FWHM of 0 implies a profile fit to the object failed. \vspace{0.5cm}}
{\normalsize
\begin{center}
\begin{tabular}{|l|l|}
\hline
Flag: & Status:
\\\hline
0 & OK
\\\hline
1 & Nearby object\\
\hline
2 & Severe overlapping with nearby object\\
\hline
3 & Object is near a saturated pixel\\
\hline
4 & Object is near a masked region\\
\hline
5 & Object is near the edge of the image\\
\hline
6 & Object is itself masked out\\
\hline
7 & Object has 0 FWHM\\
\hline
8 & Too few background pixels around object\\
\hline
9 & Object entirely overlapped by neighbors\\
\hline
10 & Routine sexcat2pstamp crashed\\
\hline 
\end{tabular}
\end{center}
\label{default}
}
\end{table}

\subsubsection{Input catalogue generation}

As described in \cite{ferry}, the first task is to generate a catalog of galaxy morphologies from real data. Note that galaxy morphologies are already provided from the UDF and, until better data become available, this time-consuming section of the pipeline need not be re-run.  However, If desired it is possible to regenerate the UDF catalogue, or indeed to generate additional catalogues.

The simulated images are based on UDF data, with photometric redshifts from \cite{coe}.  The process of getting from real data to simulated images, using modules included in the simulation pipeline software package, is as follows.  First, objects in the real data are detected and cataloged using the \emph{SExtractor} routine on the image using a specified configuration file, {\bf config.sex} an example of which can be found in the \emph{analysis} directory of the \emph{simage} package.  The cataloged objects are then decomposed into shapelets by running {\bf shex.pro}.  This program takes the real image, the \emph{SExtractor} catalog, and the desired $n_{\rm max}$ as inputs, and outputs a catalog of shapelet coefficients for the objects in the \emph{SExtractor} catalog.  We chose $n_{\rm max}=20$ for the optical band - which is sufficient to model the HST PSF.  {\bf shex.pro} also uses the shapelets focus suite of routines to optimize the $n_{\rm max}$, $\beta$, and centroid parameters.  The program also outputs flags, from 0 to 10, to tell the user how well a given object was modeled with shapelets, 0 being good and 10 signaling failure.  A list of the flags and their corresponding criteria is shown in Table 3. There is also an option in {\bf shex.pro} to remove a specified constant PSF, which can be modeled from the stars in the image.  The PSF removal from the UDF catalog is a three step process.  Firstly, the HST PSF is modeled by selecting stars in the UDF images and decomposing them into a sum of shapelet basis functions.  These stars are the best representation of the PSF contained within the image.   Secondly, the galaxy objects of an image, once also decomposed into shapelet space, have the star/PSF shapelets subtracted, thus leaving a catalog of galaxies with the HST PSF removed.  The stars in the newly created catalog are then discarded, since any PSF model that will be introduced will be formed from new simulated stellar images. 

Galaxies then have to be cross-matched across bands of data.  This is done using {\bf srcor.pro}, in the IDL Astronomy Library, to a tolerance of 1 arcsecond.  Some galaxies will appear in all bands and some will appear in a subset, but each galaxy is given an ID number and then a master catalog is created that contains the information for each unique galaxy ID, including its position, redshift, and shapelet coefficients in each band.  For the UDF, this catalog is stored as a structure called {\bf shapecat\_total\_trim.sav}, which should be located in the specified {\it data} directory alongside the necessary {\it psf} folder.  The {\bf simage} program then randomly draws galaxies (in the form of their decomposed shapelet coefficients) from this catalog when producing simulated images.

\subsubsection{Noiseless image simulation}

The pipeline would then proceed as follows:  The simulated image is scaled according to the instrument and filter throughput. These are defined by parameters \emph{throughput\_ratio} and \emph{filter\_files}.  The first specifies the throughput ratio for each band compared to that of the HST-Advanced Camera for Surveys (ACS), should it already known by the user. The second option allows you to calculate this \emph{throughput\_ratio} array by specifying an arbitrary filter curve. This \emph{filter\_files} lists the total desired instrumental throughput for each band, in the format of wavelength (\AA) vs. throughput, where 1.0 represents 100\% transmission.  The filter curve is then integrated and compared against the HST/ACS filter
curve to generate the correctly normalized array \emph{throughput\_ratio}.  In this way, it is possible to use UDF images but normalized for the
throughput of any given telescope design.  However, note that while the integral of the throughput is considered, its shape is not.  For example, the
shape of galaxies is therefore not changed by a transmission curve that peaks redward of the HST/ACS filters and should therefore enhance the bulges of galaxies.  The pipeline then reads in the specified PSF file, which remains constant throughout the simulation.

The \emph{simage} routine will then pass to \emph{simage\_assemble \_image} which reads in the UDF shapelet catalog and populates the image with either \emph{n\_gal} number of galaxies, or a pre-defined default that is calibrated to existing observational data (the HST Cosmos survey; \citealt{scov}).  At this point any specified weak lensing shear, represented by the 2D array [$\gamma_{1}$,$\gamma_{2}$], is added to each galaxy, i.e a constant value across the field (see \S3.4).  The pixel scale of the shapelet catalog galaxies are adjusted to the specified value at this stage.  A similar process is performed to populate the image with field stars, each of which are represented by the PSF.  Here the subroutine \emph{simage\_star\_magnitude\_distribution} generates a random flux level for stars in the image. The stars follow the stellar luminosity function measured at the galactic poles and tabulated in \cite{alle}.    

We note that the pipeline also has the ability to create a simulated image containing objects with analytic (e.g. de Vaucouleurs) profiles with the same size, magnitude and ellipticity distributions as the UDF shapelet catalog.  The shapelet coefficients are not used for any purpose other than the determination of these distributions.  Hence the pipeline can also use a non-shapelet method to create simulations, which can themselves be used to test shapelet based shape-measurement techniques if desired.

The output noiseless image is written in units of photons or counts per second, along with a mock \emph{SExtractor} output file with the all the objects' known input positions, sizes, magnitudes, ellipticities and star/galaxy classifications.  All files are output to the \emph{Data} directory. 

Although the galaxy catalogues are created from the UDF, the magnitude distribution of the resultant simulated images is normalized to a variety of existing galaxy surveys (described below) rather than just drawing randomly from the UDF galaxy magnitudes.  Since the decomposed shapelet objects have color representative of real galaxies, this approach can apply to all bands available to the simulation pipeline (i.e. \emph{B, V, i} and \emph{z}).  The routine {\bf num\_counts\_frac.pro} utilizes the following number (N) counts relation;
\begin{equation}
N = B*10^{(A*m)}
\end{equation}

\begin{figure}[t]
\centering
\epsfig{figure=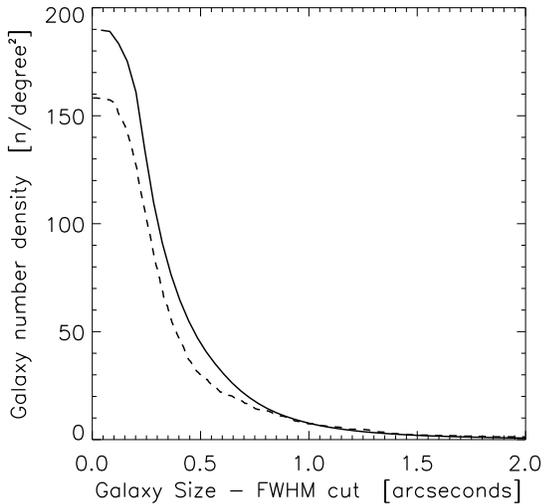,width=80mm} 
\caption{A comparison of galaxy number density vs. galaxy size for both real HST-COSMOS survey data (solid line; \citealt{Leau}) and \emph{simage} simulated COSMOS data (dashed line).  The galaxy size is measured via the SExtractor \emph{FWHM\_ IMAGE}.  The x-axis scale is the FWHM cut size - i.e.  galaxies that have a given size and above.  Simulated data used COSMOS survey parameters of 0.03 arcseconds/pixel and an exposure time of 2000 seconds.}  
\label{fig:three}
\end{figure}

\noindent where \emph{A} and \emph{B} are normalization factors, and \emph{m} is the galaxy magnitude.  The form of this expression and the values of the normalization factors A and B are taken from existing COSMOS survey analysis between magnitudes 21 and 26 \citep{Leau}.  COSMOS is the widest HST survey and as such is least affected by cosmic variance/sample size.  It is also the closest data to the future space surveys that \emph{simage} aims to model.  Below magnitude 21 and above magnitude 26 the the number counts are fit to various survey sources as compiled in \cite{Metc}.  The form is continuous at these transition points and results in a realistic number count relation for the simulated images at low, mid, and high magnitudes.

In figure \ref{fig:three} we display a comparison of galaxy sizes in the HST-COSMOS survey and those from an \emph{simage} simulated COSMOS field.  We see that the normalization of \emph{simage} to the COSMOS counts discussed above allows the pipeline to obtain a very similar sizes distribution.  We note that there are differences at the lower size limit due to \emph{simage} reproducing fewer very small galaxies in the simulation.  This is because the pipeline does not decompose the smallest galaxies into shapelets since some of them are noise (or noisy) and as such do not end up in the shapelets catalog.  

\subsubsection{Addition of noise}
 
The routine {\bf simage\_add\_noise.pro} reads in the \emph{.\,fits} format noise-free image, and writes out a noisy image, plus an inverse variance weight map. This is very fast to run, and is intentionally kept separate from the previous sections because a common task is to investigate the effect of changing the survey exposure time. In this case, the same noise-free image can be used, and {\bf simage\_add\_noise.pro} run multiple times in isolation, with different input parameters.  Specific noise features and detector effects that can be added to an image post-creation, e.g. dark current, are included within the {\bf simage\_add\_noise.pro} routine itself and are activated by selecting the corresponding flags when the routine is initially called.

Noise is then added to the image. For convenience, the noise model is calculated in two separate components: shot noise on astrophysical sources, and on the sky background. In both cases, a random distribution of uncorrelated pixel values is drawn from a Gaussian with width equal to the square root of the counts in each pixel.  The sky background level is estimated by default from that in the UDF, but can also be specified in {\bf
telescope.param}, in units of counts/second/arcsec$^2$.  By default, the constant sky background level is then subtracted, although this behavior
can be turned off via the BACKGROUND keyword. 

Additional options include the ability to add read noise, to correlate the background noise to cheaply mimic the effects of DRIZZLE, to truncate saturated pixels, and to truncate at zero any non-physical, slightly negative pixel values (arising from noise or modeling problems in the original UDF). None of these are enabled by default.  To mimic the effects of DRIZZLE, the image is convolved with a kernel similar to the drizzle drop kernel and the original pixel square. This has the effect of correlating adjacent pixels in the image, and is most noticeable in the background noise in blank areas well away from sources.

\section{Summary}

We have introduced an IDL simulation pipeline, \emph{simage}, that creates mock deep field survey images by drawing from a catalog of Hubble Ultra Deep Field galaxies.  Each galaxy in the catalog is decomposed into a set of analytic shapelet basis functions which can completely describe their morphological properties in a simple manner.  We have shown how this catalog can then be used to populate a field of any given size and resolution depending on the user's requirements.  The pipeline allows the user complete control over parameters such as exposure time, PSF type, mirror size, pixel scale, field star density, and noise, and simulates fields in the \emph{B, V, i} and \emph{z} bands.

The code also has the ability to introduce a weak lensing signal into the data, allowing the output to be used for studies into weak lensing reconstruction analysis.  It is envisioned that the code will be used as a tool for research, instrumental development, and teaching.  It is available to download as a self contained package of IDL modules at \emph{www.astro.caltech.edu/~$\sim$rjm/shapelets.}

\section*{Acknowledgments}

Thanks to Molly Peeples and Banaby Rowe.  BMD and JR acknowledges financial support through the NASA-JPL Award/Project \emph{"Weak Gravitational Lensing: The Ideal Probe of Dark Matter and Dark Energy"}.  RJM acknowledges financial support through European Union grant MIRG-CT-208994 and STFC Advanced Fellowship PP/E006450/1.  The work of BMD, JR, and AV were carried out at the Jet Propulsion Laboratory, California Institute of Technology, under contract with the National Aeronautics and Space Administration (NASA).

\section*{Appendix A: Shapelets formalism}

Shapelets come in two flavors: Cartesian shapelets are separable in \emph{x} and \emph{y}, and polar shapelets in \emph{r} and $\theta$. There is a one-to-one mapping between the two, so without loss of generality, we shall adopt whichever has the more convenient symmetries for the task at hand. The polar shapelet basis functions

\begin{eqnarray}
\chi_{n,m}(r,\theta;\beta) = 
  \frac{(-1)^{\frac{n-|m|}{2}}}{\beta^{|m|+1}} 
  \left[\frac{\left(\frac{n-|m|}{2}\right)!}{\pi \left(\frac{n+|m|}{2}\right)! }\right]^{\frac{1}{2}}
  ~\times \nonumber \\
  r^{|m|}L_{\frac{n-|m|}{2}}^{|m|} 
  \left(\frac{r^2}{\beta^2}\right)
  e^\frac{-r^{2}}{2\beta^2}
  e^{-im\theta},
\end{eqnarray}

\noindent where $L_p^q(x)$ are the Laguerre polynomials, have an overall scale size $\beta$ and are parameterized by two integers, $n$ and $m$, which are the number of oscillations in the radial and tangential directions.  The basis functions are calculated using {\bf shapelets\_chi.pro}.  Using {\bf shapelets\_decomp.pro}, a galaxy (or star) image $I(r,\theta)$ can then be decomposed into (complex) ``shapelet coefficients'' $f_{n,m}$ 

\begin{equation} \label{eqn:lindecompp}
f_{n,m} = \iint_{\mathbb{R}} I(r,\theta) ~\chi_{n,m}(r,\theta;\beta)~r~{\mathrm d}r {\mathrm d}\theta ~,
\end{equation}
so that the (wholly real) image can be reconstructed, using {\bf shapelets\_recomp.pro} as
\begin{equation} \label{eqn:sseriesp}
I(r,\theta) = \sum_{n=0}^\infty \sum_{m=-n}^{n} f_{n,m} \chi_{n,m}(r,\theta;\beta) ~.
\end{equation}

In practice, it is necessary to truncate the expansion at some maximum value of $n$. Figure ~\ref{fig:one} shows an example galaxy image and its reconstructed counterpart using shapelets up to order $n_{\mathrm max}=20$.  It can be seen that the model easily captures the major features of the original galaxy.

In shapelet representation, convolution between two images (such as a galaxy and a telescope's Point Spread Function) is
simply a matrix multiplication of their $f_{n,m}$ coefficient arrays \citep{refr}. In the code, this is implemented via
{\bf shapelets\_convolve.pro}. It is also possible to perform a deconvolution by inverting the PSF matrix; this is
incorporated within {\bf shapelets\_decomp.pro}.

While previous operations were performed in polar shapelets since the functions are separable in $r$ and $\theta$ (rendering many operations more intuitive), pixillation can be performed most easily by switching to Cartesian shapelets, then switching back.  A closed form for the integrals of Cartesian shapelet basis functions over rectangular pixels is given in \S4.3 of \cite{mass}, and is enabled by default in {\bf shapelets\_chi.pro}.

We shall use several transformations of the galaxy images, first to randomize their appearance in the final simulation,
and then to impose a gravitational lensing signal. In shapelet space, galaxies can be easily rotated by adjusting the
phase of their (complex) coefficients $f_{n,m}$, or reflected in the $x$-axis by taking their complex conjugates.
A weak gravitational lensing shear signal $\gamma$ can be applied to first order by mixing adjacent coefficients according to the mixing matrix

\begin{eqnarray} \label{eqn:opshear}
\widehat{1}+\widehat{\shear}(\gamma):f_{n,m} \rightarrow f_{n,m}^\prime = f_{n,m} ~~~~~~~~~~~~~~~~~~~~~~~ \\
+ \frac{\gamma}{4}\left\{ \sqrt{(n+m)(n+m-2)}~f_{n-2,m-2} \right. ~~~~~~~~~ \nonumber \\
- \left. \sqrt{(n-m+2)(n-m+4)}~f_{n+2,m-2} \right\} \nonumber \\
+ \frac{\gamma^*}{4}\left\{ \sqrt{(n-m)(n-m-2)}~f_{n-2,m+2} \right. ~~~~~~~~~ \nonumber \\
- \left. \sqrt{(n+m+2)(n+m+4)}~f_{n+2,m+2} \right\} \nonumber
\end{eqnarray}

\noindent as described in \S2.3 of \cite{mass2}, which also provides similar operations for flexion.  In the above, $\widehat{1}$ corresponds to the identity operator, while $\widehat{\shear}$ corresponds the to shear operator.
Routines to implement such operations in practice are located in the {\bf shapelets/operations/} subdirectory.

The above prescription for shear is only accurate to order $\gamma$. This will be insufficiently accurate for very
high precision work, or if the gravitational lensing signal is particularly large. A new innovation for \emph{simage}
is that this transformation can now be generalized to include higher order $\gamma^2$, $\gamma^3$, etc.\ terms.
This is achieved mathematically by exponentiating the operation \citep{bern}. For a practical implementation to 
fourth order, note that the first four terms in an exponential expansion

\begin{eqnarray}
\widehat{1}+\widehat{\shear}+\frac{\widehat{\shear}^2}{2!}+\frac{\widehat{\shear}^3}{3!}+\frac{\widehat{\shear}^4}{4!} = \frac{3}{8} + ~~~~~~~~~~~~~~~~~~~~~~~\nonumber \\
  \frac{\widehat{1}+\widehat{\shear}}{3} + \frac{(\widehat{1}+\widehat{\shear})^2}{4} + \frac{(\widehat{1}+\widehat{\shear})^4}{24} 
\end{eqnarray}

\noindent where $\widehat{\shear}$ here is the shear operator but could equally be replaced by any other.
To perform this on a shapelet model we simply need to apply the linear mapping~(\ref{eqn:opshear}) four times, 
recording the new coefficients $f^\prime_{n,m}$ at each stage, and add them to the original coefficients in the 
ratio $\frac{3}{8}:\frac{1}{3}:\frac{1}{4}:0:\frac{1}{24}$. This behavior is controlled via the ORDER keyword in
{\bf shapelets\_shear.pro}, and is set to 4 by default.

Note that, as discussed in BJ02, there is a somewhat arbitrary choice for these higher order terms, which can be
changed depending on the required definition of shear. The expansion above changes an intrinsically circular
source into an ellipse with major and minor axes $a$ and $b$ via a distortion
$\delta\equiv(a^2-b^2)/(a^2+b^2)$.  A ``conformal shear'', $\nu=\mathrm{arctanh}(\delta)$, produces a slightly
different ratio of major and minor axes, but can be achieved by simply adjusting the input shear.   A fourth-order
implementation of a conformal shear in shapelet space perfectly matches the real-space transformation of highly
oversampled images within a computer's numerical precision (B.\ Rowe, priv.\ comm. 2008). It is therefore not just
faster, but should be accurate within 1\% for shears up to $\gamma\approx0.47$ \citep{gold}. For a typical
cosmological gravitational lensing signal of a few percent, applying only a first order shapelet-based shear
yields pixel values in the final image that are incorrect at a level of approximately one part in $10^{-3}$, and
changing from $\delta$ to $\nu$ yields differences of around one part in $10^{-5}$.  Because of this, the \emph{simage} pipeline routines such {\bf simage\_make\_analytic\_object.pro}, uses the conformal shear, $\nu$, when called to include a shear signal.


\begin{thebibliography}{99}
\bibitem[\protect\citeauthoryear{Allen}{2000}]{alle} Allen, 2000, ``Allen's Astrophysical Quantities'', Cox A. (Ed), The Athlone Press, London, U.K., 4th Edition
\bibitem[\protect\citeauthoryear{Beckwith et al.}{2006}]{beck} Beckwith S. V. W. et al., 2006, AJ, 132, 1729
\bibitem[\protect\citeauthoryear{Bernstein \& Jarvis}{2002}]{bern} Bernstein G., \& Jarvis M., 2002, AJ, 123, 583
\bibitem[\protect\citeauthoryear{Bridle et al.}{2009}]{brid2} Bridle S., et al., 2009, arXiv:0908.0945
\bibitem[\protect\citeauthoryear{Bridle et al.}{2008}]{brid} Bridle S., et al., 2008, AnApS, 3, 6
\bibitem[\protect\citeauthoryear{Coe et al.}{2006}]{coe} Coe D., Ben'tez N., S\'anchez S., Jee M., Bouwens R., Ford H., 2006, AJ, 132, 926.  http://adcam.pha.jhu.edu/~coe/UDF/
\bibitem[\protect\citeauthoryear{Erben et al.}{2001}]{erbe} Erben T., van Waerbeke L., Bertin E., Mellier Y., Schneider P., 2001, A\&A, 366, 717
\bibitem[\protect\citeauthoryear{Ferry et al.}{2008}]{ferry} Ferry M., Rhodes J., Massey R., White M., Coe D., Mobasher B., 2008, APh, 30, 65
\bibitem[\protect\citeauthoryear{High et al.}{2007}]{high} High F. W., Rhodes J., Massey R., Ellis R., 2007, PASP, 119, 1295
\bibitem[\protect\citeauthoryear{Ivezic}{2007}]{ivez} Ivezic Z., 2007, AAS, 210, 6605
\bibitem[\protect\citeauthoryear{Jelinsky}{2006}]{jeli} Jelinsky P., 2006, AAS, 209, 9809 
\bibitem[\protect\citeauthoryear{Leauthaud et al.}{2007}]{Leau} Leauthaud A., et al., 2007, ApJS, 172, 219
\bibitem[\protect\citeauthoryear{Refregier et al.}{2003}]{refr} Refregier A., 2003, MNRAS, 338,35
\bibitem[\protect\citeauthoryear{Refregier \& Bacon}{2003}]{refr2} Refregier A., \& Bacon D., 2003, MNRAS, 338, 48
\bibitem[\protect\citeauthoryear{Refregier et al.}{2010}]{refr3} Refregier A., \& The Euclid Imaging Consortium, et al., 2010, arXiv:1001.0061
\bibitem[\protect\citeauthoryear{Meneghetti et al.}{2008}]{Mene1} Meneghetti M., Melchior P., Grazian A., De Lucia G., Dolag K., Bartelmann M., Heymans C., Moscardini L., Radovich M., 2008, A\&A, 482, 403
\bibitem[\protect\citeauthoryear{Metcalfe et al.}{2001}]{Metc} Metcalfe N., Shanks T., Campos A., McCracken H. J., \& Fong R., 2001, MNRAS, 323, 795.
\bibitem[\protect\citeauthoryear{Massey et al.}{2004}]{simage1} Massey R., Refregier A., Conselice C., Bacon D., 2004, MNRAS, 348, 214
\bibitem[\protect\citeauthoryear{Massey \& Refregier}{2005}]{mass} Massey R., \& Refregier A., 2005, MNRAS, 363, 197 
\bibitem[\protect\citeauthoryear{Massey et al.}{2007a}]{mass2} Massey R., Rowe B., Refregier A., Bacon D., Berge J., 2007a, MNRAS, 380, 229
\bibitem[\protect\citeauthoryear{Massey et al.}{2007b}]{step2} Massey R., et al., 2007b, MNRAS, 376, 13
\bibitem[\protect\citeauthoryear{Massey \& Goldberg}{2008}]{gold} Massey R., \& Goldberg D., 2008, ApJL, 673, 111
\bibitem[\protect\citeauthoryear{Schrabback et al.}{2009}]{schr} Schrabback T., et al., 2009, arXiv:0911.0053
\bibitem[\protect\citeauthoryear{Scoville et al.}{2007}]{scov} Scoville N., et al., 2007, ApJS, 172, 1
\bibitem[\protect\citeauthoryear{Vanderveld et al.}{2010}]{vand} Vanderveld R. A., et al., 2010, in prep.
\end{thebibliography}
\end{document}